\documentclass[aps,pre,groupedaddress,showpacs,preprint]{revtex4}
\usepackage{graphicx}% Include figure files
\usepackage[dvips]{epsfig}% Include figure files
\usepackage{dcolumn}% Align table columns on decimal point
\usepackage{subfigure}%
\usepackage{bm}% bold math
\usepackage{amssymb}
\usepackage{amsmath}
\usepackage{color}

\begin{document}
\title{Origins of scaling relations in nonequilibrium growth}

\author{Carlos Escudero$^\dag$ and Elka Korutcheva$^\ddag$$^,$\footnote{Also at G. Nadjakov Inst. of Solid State Physics, Bulgarian Academy
of Sciences, 1784 Sofia, Bulgaria.}}

\affiliation{$\dag$ Departamento de Matem\'aticas \& \\
ICMAT (CSIC-UAM-UC3M-UCM), \\
Universidad Aut\'onoma de Madrid, \\
Ciudad Universitaria de Cantoblanco, E-28049 Madrid, Spain
\\
$\ddag$ Departamento de F\'{\i}sica Fundamental, \\
Universidad Nacional de Educaci\'{o}n a Distancia, C/ Senda del
Rey 9, E-28040 Madrid, Spain}

\begin{abstract}
Scaling and hyperscaling laws provide exact relations among critical
exponents describing the behavior of a system at criticality. For
nonequilibrium growth models with a conserved drift there exist few
of them. One such relation is $\alpha +z=4$, found to be inexact in
a renormalization group calculation for several classical models in
this field. Herein we focus on the two-dimensional case and show
that it is possible to construct conserved surface growth equations
for which the relation $\alpha +z=4$ is exact in the renormalization
group sense. We explain the presence of this scaling law in terms of
the existence of geometric principles dominating the dynamics.
\end{abstract}


\maketitle

\section{Introduction}
\label{introduction}

One of the great successes of equilibrium
statistical mechanics is the classification of critical phenomena
into universality classes. These are characterized by sets of
critical exponents which describe the system long range physics in
the critical point. It is precisely the infrared behavior, which
is independent of the microscopic details of the particular
system, which lies at the basis of universality.

The numerical values of the critical exponents can be calculated
exactly in a number of fortunate cases, otherwise they have to be
calculated approximately using some perturbative technique, like
notably the renormalization group, or by means of numerical
simulations. Scaling and hyperscaling laws provide exact relations
among critical exponents and consequently both are a deep
expression of the physics and provide a useful calculational tool.

Nonequilibrium statistical mechanics has been built in many
situations as an extension of its successful equilibrium
counterpart. The classification in terms of universality classes
and the tools for computing critical exponents have been adapted
to systems out of equilibrium, such as absorbing state
transitions~\cite{henkel} and stochastic growth~\cite{barabasi}.
This last field is sometimes regarded as paradigmatic within
nonequilibrium statistical mechanics. The reason for this is the
ubiquity of certain universality classes which were originally
discovered in this context and were subsequently found in
different areas of physics. A particularly relevant example of
this is the Kardar-Parisi-Zhang (KPZ) equation~\cite{kpz}, which
scaling behavior has been related to phenomena as diverse as far
from equilibrium growth, turbulence and directed polymers.

The KPZ and other nonequilibrium growth equations can be
characterized by three scaling exponents: the roughness exponent
$\alpha$ characterizing the interface morphology, the dynamic
exponent $z$ which specifies the velocity at which correlations
propagate and the growth exponent $\beta$ describing its short
time dynamics~\cite{barabasi}. One of the most important relations
among these exponents is the scaling relation $\alpha=\beta z$,
which is fulfilled in a large number of cases. Together with it
there exist other scaling and hyperscaling relations describing
fundamental characteristics of different physical situations.

This work is devoted to the study of stochastic partial
differential equations with a conserved drift. By this we mean
that the deterministic part is conservative or, in other words,
the spatial average of the deterministic part of the equation
vanishes provided suitable boundary conditions are used.

In the following we will examine one of this scaling relations
already found in the literature, and we will investigate its
possible physical origin. In this work we concentrate on the
two-dimensional situation. The paper is organized as follows: in
Section II we present some scaling relations for nonequilibrium
models, in Section III we give a short geometric derivation of a
family of conserved growth equations we will be interested in.
Section IV is devoted to the renormalization group analysis of
these and other related equations. In the last two Sections we
discuss the possible origin of scaling relations for the different
models and draw our conclusions.

\section{Scaling relations}

The KPZ equation reads~\cite{kpz}
\begin{equation}
\partial_t h = \nu \nabla^2 h + \frac{\lambda}{2} (\nabla h)^2 +
\xi(\vec{r},t),
\end{equation}
where $\xi$ is a zero mean Gaussian noise which correlation is
given by $\langle \xi(\vec{r},t)\xi(\vec{r} \, ',t') \rangle = D
\delta(\vec{r}-\vec{r} \, ') \delta(t-t')$. If one performs the
dilatation $\vec{r} \to b \, \vec{r}$, $t \to b^z \, t$ and $h \to
b^\alpha \, h$ the nonlinearity becomes scale invariant provided
$\alpha + z =2$. This scaling relation is actually fulfilled by
the KPZ equation in any dimension and its exactness is usually
attributed to its Galilean invariance property~\cite{fns,medina}.
However the role of Galilean invariance was put into question in a
related nonequilibrium model~\cite{mccomb,hochberg,hochberg2}. In
particular, in these works it is shown that neither Galilean
invariance nor extended Galilean invariance contribute to vertex
non-renormalization beyond the zeroth mode. Also, the same scaling
relation $\alpha + z =2$ was found in numerical
schemes~\cite{wio,wior,wiop} and stochastic
equations~\cite{nicoli,nicoli2} which do not explicitly fulfill
the Galilean invariance symmetry.

Together with the KPZ equation, in this field one finds the
Sun-Guo-Grant (SGG)~\cite{sun} and Villain-Lai-Das Sarma
(VLDS)~\cite{villain,lai} equations, which read
\begin{equation}
\label{ssg}
\partial_t h = -\nu \nabla^4 h + \lambda \nabla^2 (\nabla h)^2 +
\xi^{(1,2)}(\vec{r},t),
\end{equation}
where the noise $\xi^{(1)}$ corresponds to the SGG equation and
$\xi^{(2)}$ to the VLDS one.

In both cases the noise is a zero
mean stochastic Gaussian process and the respective correlations
are:
\begin{equation}\langle \xi^{(1)}(\vec{r},t)\xi^{(1)}(\vec{r} \, ',t') \rangle = -D \nabla^2
\delta(\vec{r}-\vec{r} \, ') \delta(t-t'),
\end{equation} and
\begin{equation}
\langle \xi^{(2)}(\vec{r},t)\xi^{(2)}(\vec{r} \, ',t') \rangle = D
\delta(\vec{r}-\vec{r} \, ') \delta(t-t').
\end{equation}
For these equations the nonlinearity becomes scale invariant to
the dilatation transformation provided the exponents fulfill the
scaling relation $\alpha + z =4$. This relation was found to be
fulfilled at one loop order renormalization group calculations and
it was conjectured to be exact due to the existence of a
functional Galilean invariance symmetry~\cite{sun}. However,
in~\cite{janssen}, Janssen showed that the mathematics leading to
this symmetry was ill-posed. Also, by means of a renormalization
group calculation, he showed that this relation does not hold at
two loops~\cite{janssen}. The correction to the scaling relation
he found is however small and presumably not detectable in
simulations or experiments.

We also would like to mention here that this correction has not
been found in the results obtained by means of other methods, such
that the self-consistent expansion~\cite{sce1,sce2,sce3}, when
applied to these equations~\cite{katzav1}. This approximation
scheme has proved itself very useful in finding exponent values in
some cases in which the renormalization group fails, like for
instance for the nonlocal counterpart of some stochastic growth
equations~\cite{katzav2,katzav3}. On the other hand,
renormalization group analyses have been able to yield exact
results for some stochastic growth equations~\cite{frey1,frey2}.
We will restrict ourselves to the renormalization group analysis
of local equations and leave different approaches for a more
comprehensive study. As will be evident in the following, this
analysis has a clear interpretation in the cases under
consideration.

In the next section we will show that it is possible to find
models for which the scaling relation $\alpha +z=4$ is exact to
all orders in a renormalization group calculation.

\section{Geometric derivation of a stochastic growth equation}

We now briefly review the geometric derivation of a stochastic
growth equation~\cite{marsili}. We concentrate on $d=2$ and assume
the following variational approach for the dynamics
\begin{equation}
\label{principle} \frac{\partial h}{\partial t}= \sqrt{1 + (\nabla
h)^2} \left( - \frac{\delta \mathcal{I}}{\delta h} + \xi^{(1,2)}
\right),
\end{equation}
where we have added the noise term $\xi^{(1,2)}$ and $\mathcal{I}$
is a nonequilibrium potential. We assume it can be expressed as a
function of the surface mean curvature only
\begin{equation}
\label{potential} \mathcal{I}= \int f(H) \, \sqrt{1 + (\nabla
h)^2} \,\, d\vec{r},
\end{equation}
where $H$ denotes the mean curvature, $f$ is an unknown function
of $H$ and whenever we are in two dimensions $\vec{r}=(x,y)$. The
presence of the square root terms in Eqs.~(\ref{principle}) and
(\ref{potential}) describes growth along the normal to the
surface.

We will further assume that
function $f$ can be expanded in a power series
\begin{equation}
\label{expansion}
f(H)= K_0+K_1 H + \frac{K_2}{2} H^2 + \cdots ,
\end{equation}
of which only the zeroth, first, and second order terms will be of
relevance at large scales.

The result of the minimization of the potential~(\ref{potential})
leads to the equation (see section~III.A.3 in~\cite{marsili})
\begin{eqnarray}
\nonumber
\partial_t h = \mu \nabla^2 h &+& \lambda \left[ (\partial_{xx}h)(\partial_{yy}h)-(\partial_{xy}h)^2
\right] \\
&-& \nu \nabla^4 h + \xi^{(1,2)},
\label{monge}
\end{eqnarray}
to leading order in the small gradient expansion, which assumes
$|\nabla h| \ll 1$. Here $\mu=K_0$, $\lambda=2K_1$ and $\nu=K_2$.
We note that this equation can be expressed as the divergence of a
current $\partial_t h = \nabla \cdot j + \xi^{(1,2)}$ for $j = \mu
\nabla h + \frac{\lambda}{2} \, \mathrm{cof}(D^2 h) \cdot \nabla h
-\nu \nabla \nabla^2 h$ provided some regularity conditions on $h$
are assumed~\cite{muller}, and where $\mathrm{cof}(D^2 h)$ is the
cofactor matrix of the Hessian matrix (see
Eq.~(\ref{hessianmatrix}) below).

There is another way of finding the first two terms of this
equation. The Hessian matrix
\begin{equation}\label{hessianmatrix}
(D^2 h)=
\left( \begin{array}{lr} \partial_{xx} h \, & \, \partial_{xy} h \\
\partial_{yx} \, h & \, \partial_{yy} h
\end{array} \right)
\end{equation}
encodes all the information about the convexity and concavity of
the interface. This matrix has exactly two tensorial invariants:
its trace and its determinant. These are in fact the first two
terms of our equation. A related viewpoint was adopted
in~\cite{hanggi} in the derivation of a model for amorphous thin
film growth.

The term proportional to the Laplacian is associated to the effect
of gravity and desorption in the framework of nonequilibrium
growth. These effects are negligible in molecular beam epitaxy
growth, and this is precisely the context in which the VLDS
equation was derived~\cite{villain,lai}. From a theoretical point
of view, the presence of this term in Eq.~(\ref{monge}) makes the
system evolve towards the linear Edwards-Wilkinson fixed point. So
theoretically it is more interesting to neglect this term and this
way find a genuine nonlinear dynamics. Setting $\mu=0$ in
Eq.~(\ref{monge}) yields
\begin{equation}\label{mampere}
\partial_t h = \lambda \left[ (\partial_{xx}h)(\partial_{yy}h)-(\partial_{xy}h)^2
\right] - \nu \nabla^4 h + \xi^{(1,2)}.
\end{equation}
This equation can be considered as an alternative to both SSG and
VLDS equations~\cite{escudero}. Indeed, it is identical to both
except for the nonlinearity, that shares the same dimensional
analysis properties and the fact that it is conserved.

\begin{figure}[h]
\begin{center}
\includegraphics[scale=0.5]{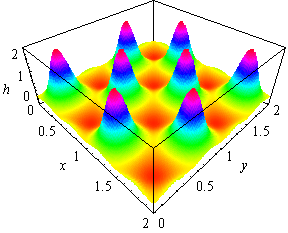}
\caption{Numerical solution of Eq.~(\ref{mampere}) in the
deterministic limit $D=0$. The values of the parameters are
$\lambda=1$ and $\nu=0.1$. We have assumed periodic boundary
conditions and the symmetric initial condition $h_0=\sin(2\pi x)
\sin(2 \pi y)$. One sees that holes disappear and mounds form as a
consequence of the action of the nonlinearity. Red and orange
colors denote negative values of $h$.} \label{mounds}
\end{center}
\end{figure}

In order to clarify the meaning of the nonlinear term in
Eq.~(\ref{mampere}) we numerically solve this equation in the
deterministic limit $D=0$, see Fig.~\ref{mounds}. We see that
holes disappear and mounds form as a consequence of the action of
the quadratic nonlinearity. This phenomenology can be explained by
means of a simple geometric argument. The surface Gaussian
curvature is given in the small gradient limit by $\mathcal{K}
\approx h_{xx}h_{yy}-h_{xy}^2$, so the dynamics favors the growth
of those parts with a positive Gaussian curvature, which is
precisely the observed effect.

As a final remark, we note that the growth of perturbations with a
defined wavelength may be achieved by means of considering the
full Eq.~(\ref{monge}) with $\mu <0$. The effect of this type of
linear instability on the VLDS equation has already been
considered~\cite{vvedensky}. It could be possible to perform a
renormalization group analysis in this case as well. One can use
the studies on the noisy Kuramoto-Sivashinsky equation which use
this technique~\cite{rodolfo,hu} as a benchmark to this purpose.

\section{Renormalization group analysis}

Herein we will analyze the equations derived in the previous
section as well as some possible variants. Our main tool will be
the dynamic renormalization group (RG) as developed
in~\cite{fns,medina}.

\subsection{Methodology}\label{method}

As a first step we will explain how to implement the RG analysis
in our current situation. Due to the particular form of the
nonlinearity, the present analysis requires considering some
subtleties which are not present in the case of more studied
nonlinear terms.

We begin with the equation
\begin{equation}\label{ssgvldsma}
\partial_t h = \lambda (h_{xx}h_{yy}-h_{xy}^2) -\nu \nabla^4 h +
\xi(\vec{r},t),
\end{equation}
forced by a white Gaussian noise characterized by its two first
moments:
\begin{eqnarray}
\langle \xi(\vec{r},t) \rangle &=& 0, \\
\langle \xi(\vec{r},t) \xi(\vec{r} \, ',t') \rangle &=& D
\delta(\vec{r}-\vec{r} \, ') \delta(t-t').
\end{eqnarray}
Next we Fourier transform this equation to find
\begin{eqnarray}
\nonumber
h(\vec{k},\omega)=h_0(\vec{k},\omega) + \lambda \, G_0(\vec{k},\omega) \times \\
\nonumber \int \left[ \frac{1}{2} k_x^2 q_y^2 + \frac{1}{2} k_y^2
q_x^2 -k_x k_y q_x q_y \right] \\
\times \, h (\vec{k}-\vec{q},\omega-\Omega) \, h(\vec{q},\Omega)
\, \frac{d\Omega \, d\vec{q}}{(2 \pi)^3}, \label{fourieramper}
\end{eqnarray}
where
\begin{eqnarray}
h_0(\vec{k},\omega) = G_0(\vec{k},\omega) \, \tilde{\xi}(\vec{k},\omega), \\
G_0(\vec{k},\omega) = (-i \omega +  \nu k^4)^{-1},
\end{eqnarray}
$\tilde{\xi}$ is the Fourier transformed white noise,
$k=|\vec{k}|$ and $\vec{k}=(k_x,k_y)$ (the same holds for
$\vec{q}$). This equation is to be solved iteratively in the
vertex. The resulting Feynman diagrams are identical to the ones
represented in~\cite{medina} and~\cite{sun}. The dynamic
renormalization group technique~\cite{fns} calculates the
intermediate values of the renormalized parameters by integrating
out fast modes in the momentum shell $e^{-\ell} \Lambda \le k \le
\Lambda$, where $\Lambda$ is the momentum cutoff and $\ell$
parameterizes the change across different scales. The remaining
slow modes ($k < \Lambda$) are restored to full momentum space by
means of rescaling space and time $\vec{k} \, '=e^{-\ell} \,
\vec{k}$, $\omega' = e^{-z \, \ell} \, \omega$, and field $h'=
e^{\alpha \, \ell} \, h$. The scaled field $h'$ obeys
Eq.~(\ref{fourieramper}) as well provided the renormalized
coefficients satisfy the renormalization group flow equations,
that we will calculate in the following.

In the present case the noise amplitude is $D$, the propagator is
\begin{equation}
\frac{1}{-i \omega + \nu k^4},
\end{equation}
and the vertex is
\begin{equation}\label{vertex}
\lambda \int \left( \frac{1}{2} k_x^2 q_y^2 + \frac{1}{2} k_y^2
q_x^2 -k_x k_y q_x q_y \right) = \lambda \int \frac{1}{2} \left(
\vec{k} \cdot \vec{q} \, ^\perp \right)^2,
\end{equation}
where $\vec{q} \, ^\perp =(q_y,-q_x)$. The last equality is
advantageous because it will allow us to integrate over the angle
formed by $\vec{k}$ and $\vec{q} \, ^\perp$. If this equality were
not present we would have to carry out a more complicated
calculation reminiscent to that described in the Appendix. Note that
here we are explicitly using we are in $d=2$. In two dimensions the
situation is particularly simple because for any given non-zero
vector there are just two vectors of a given modulus perpendicular
to it. Our selection is arbitrary: choosing the opposite possibility
for $\vec{q} \, ^\perp$ does not modify the result as the scalar
product on the right hand side of Eq.~(\ref{vertex}) is squared.
Designing the optimal vertex form for a similar model built in $d>2$
is a more technical issue.

The first diagrammatic contribution to propagator renormalization
is
\begin{eqnarray}
\nonumber \frac{\lambda^2}{8 \pi^3} \frac{1}{(-i \omega + \nu
k^4)^2} \int \left[ \frac{1}{2} (k_x-q_x)^2 q_y^2 + \frac{1}{2}
(k_y-q_y)^2
q_x^2 -(k_x-q_x) (k_y-q_y) q_x q_y \right] \times \\
\left( \frac{1}{2} k_x^2 q_y^2 + \frac{1}{2} k_y^2 q_x^2 -k_x k_y
q_x q_y \right) \frac{1}{-i \Omega + \nu q^4} \frac{1}{i \Omega +
\nu q^4} \frac{D}{-i (\omega - \Omega) + \nu |\vec{k}-\vec{q}|^4}
\, d\vec{q} \,\, d\Omega.
\end{eqnarray}
It is important to realize that we have the simplification
\begin{equation}\label{simplification}
\frac{1}{2} (k_x-q_x)^2 q_y^2 + \frac{1}{2} (k_y-q_y)^2 q_x^2
-(k_x-q_x) (k_y-q_y) q_x q_y = \frac{1}{2} k_x^2 q_y^2 +
\frac{1}{2} k_y^2 q_x^2 -k_x k_y q_x q_y.
\end{equation}
Now using Eq.~(\ref{vertex}) we reduce this diagrammatic
contribution to
\begin{equation}
\frac{\lambda^2}{32 \pi^3} \frac{D}{\nu^2 k^8} \int \frac{k^4 q^4
\cos^4(\theta)}{(-i \Omega + \nu q^4)(i \Omega + \nu q^4)^2} q \,
dq \, d\theta,
\end{equation}
where $\theta$ is the angle formed by $\vec{k}$ and $\vec{q} \,
^\perp$, and we have carried out the limit $k,\omega \to 0$. We note
this contribution is relevant because the integrand is $O(k^4)$ and
the propagator renormalizes at this same order. Diagrammatically
this one loop order is identical to the one represented, for
instance, in figure 2(a) in~\cite{medina} and figure 1(b)
in~\cite{sun}.

The first diagrammatic contribution to noise renormalization is
\begin{equation}
\frac{1}{8 \pi^3} \frac{\lambda^2}{\nu^2 k^8} \int
\frac{D^2}{(\nu^2 q^8 + \Omega^2)^2} \left( \frac{1}{2} k_x^2
q_y^2 + \frac{1}{2} k_y^2 q_x^2 -k_x k_y q_x q_y \right)^2 \,
d\vec{q} \,\, d\Omega,
\end{equation}
in the limit $k,\omega \to 0$. This contribution is irrelevant
because the integrand is $O(k^4)$. The corresponding diagrams at
one loop are the same as those represented, for instance, in
figure 2(b) in~\cite{medina} and figure 1(c) in~\cite{sun}.

The first diagrammatic contribution to vertex renormalization is
irrelevant for exactly the same reason. Diagrammatically this one
loop order is identical to the one represented, for instance, in
figure 2(c) in~\cite{medina} and figure 1(d) in~\cite{sun}. In
this case all diagrams have three external legs: one corresponds
to an ongoing $\vec{k}_1$ momentum and the other two to outgoing
$\vec{k}_2$ and $\vec{k}_1-\vec{k}_2$ momenta. Consider for
instance the intermediate vertex corresponding to an ongoing leg
$\vec{k}_1$ and two outgoing legs $\vec{q}$ and $\vec{k}_1 -
\vec{q}$:
\begin{equation}
\frac{1}{2} (k_{1x}-q_x)^2 q_y^2 + \frac{1}{2} (k_{1y}-q_y)^2
q_x^2 -(k_{1x}-q_x) (k_{1y}-q_y) q_x q_y
\end{equation}
which is of order $O(k^2)$. The other two intermediate vertices
are trivially of order $O(k^2)$, so the order of this diagrammatic
contribution is $O(k^6)$. Automatically this implies this
contribution is irrelevant in the hydrodynamic limit $k \to 0$,
because the vertex renormalizes at order $O(k^4)$.

The renormalization group flow equations then read
\begin{eqnarray}
\label{rgf1} \frac{d \nu}{d \ell} &=& \nu \left[ z-4-\frac{3}{64
\pi} \frac{\lambda^2 D}{\nu^3} \right], \\ \label{rgf2}
\frac{dD}{d\ell} &=& D \left[ z-2-2\alpha \right], \\
\label{rgf3} \frac{d \lambda}{d\ell} &=& \lambda \left[ z+\alpha
-4 \right].
\end{eqnarray}
These equations yield the following values of the critical
exponents: $\alpha=2/3$ and $z=10/3$. The exponents are exactly
the ones found for the VLDS equation at one loop order. The
difference in this case is that the result is exact. We note that
noise does not renormalize at any order in the loop expansion
yielding the exactness of Eq.~(\ref{rgf2}). This is a well known
fact due to the conserved character of the drift of the stochastic
growth equation. In the present case, Eq.~(\ref{rgf3}) is exact
too due to the fact that the vertex does not renormalize at any
order in the loop expansion. All diagrams contributing to vertex
renormalization at any order are irrelevant in exactly the same
way we have shown herein for the one loop order diagrams.

As we have seen, the exactness of the scaling relation $\alpha + z
=4$ holds in this case. The current situation is genuinely
different from the one concerning both SGG and VLDS equations. In
that case vertex non-renormalization was fulfilled at one loop due
to the cancellation of all diagrams contributing at this order.
Janssen showed that the cancellation of diagrams contributing to
vertex renormalization at two loops is not exact, what leads to
the inexactness of the scaling relation~\cite{janssen}. In the
present case this argument does not apply because there is no
cancellation of diagrams: instead all diagrams contributing to
vertex renormalization are irrelevant. So the reason underlying
the exactness of the scaling relation is fundamentally different.

It is important to note we have used a symmetrized vertex given by
Eq.~(\ref{vertex}) after using the simplification
Eq.~(\ref{simplification}). This choice is by no means unique, as
we could have chosen a different vertex form, such as
\begin{equation}\label{asymmtric}
\lambda \int \left[ (k_x-q_x)^2 q_y^2 -(k_x-q_x) (k_y-q_y) q_x q_y
\right],
\end{equation}
or an infinite number of different possibilities. Among all of
them, the only one which is symmetric under a simultaneous change
of coordinates $(k_x \longleftrightarrow k_y, q_x
\longleftrightarrow q_y)$ is Eq.~(\ref{vertex}). Contrary to what
we have shown herein, using an asymmetric form results in the
dependence of the RG flow equations on the polar angle in Fourier
space $\phi$. Indeed, the substitution $k_x=k \, \cos(\phi)$ and
$k_y = k \, \sin(\phi)$ in Eq.~(\ref{asymmtric}) does not lead to
a vertex form independent of $\phi$. Formally proceeding, this
dependence is translated to the flow equations. This leads to a
perfectly valid result whenever the resulting exponents are
independent of this angle, see~\cite{escudero}. However, in the
general case, some dependence on this angle may arise. The
presence of this angle in the RG flow equations signals the
generation of an integral operator in the stochastic growth
equation when its coefficients become renormalized. This
consequently leads to a more complicated mathematical treatment,
as it implies the necessity of restarting the RG procedure
including the newly generated operators. In order to avoid this,
the symmetric form Eq.~(\ref{vertex}) must be used. We discuss in
Appendix~\ref{appasymm} the different results which may arise from
the RG analysis employing asymmetric vertex forms.

\subsection{Results for different models}
\label{differentmodels}

Now we will explore different stochastic growth equations with the
methodology explained in the previous section. Our goal is getting
a deeper understanding of the properties of the nonlinearity we
have considered so far or variants of it. In particular, we will
show that vertex non-renormalization is present in those cases in
which noise renormalization is present.

We start with a variation of Eq.~(\ref{ssgvldsma})
\begin{equation}
\partial_t h = \lambda (h_{xx}h_{yy}-h_{xy}^2) -\nu \nabla^4 h +
\xi(\vec{r},t),
\end{equation}
where now the noise is conserved, i.~e. it is white and Gaussian
and its two first moments are given by
\begin{eqnarray}\label{cnoise1}
\langle \xi(\vec{r},t) \rangle &=& 0, \\
\langle \xi(\vec{r},t) \xi(\vec{r} \, ',t') \rangle &=& -D
\nabla^2 \delta(\vec{r}-\vec{r} \, ') \delta(t-t').\label{cnoise2}
\end{eqnarray}
As we have already outlined, this equation can be considered as an
analogue of the SSG one. The renormalization group flow equations
can be obtained following analogous arguments to those of the
previous section and read
\begin{eqnarray}\label{rgf11}
\frac{d \nu}{d \ell} &=& \nu \left[ z-4-\frac{3}{64 \pi}
\frac{\lambda^2 D}{\nu^3} \right], \\ \label{rgf22}
\frac{dD}{d\ell} &=& D \left[ z-4-2\alpha \right], \\
\label{rgf33} \frac{d \lambda}{d\ell} &=& \lambda \left[ z+\alpha
-4 \right].
\end{eqnarray}
From here we get the critical exponents $\alpha=0$ and $z=4$, this
is, the SSG universality class. We emphasize that the last two RG
flow equations are exact, i.~e. both noise and vertex do not
renormalize at any order in the loop expansion, for exactly the
same reasons as in the previous section.

As the next example we consider the following equation
\begin{equation}\label{sixor}
\partial_t h = \lambda (h_{xx}h_{yy}-h_{xy}^2) +\nu \nabla^6 h +
\xi(\vec{r},t),
\end{equation}
where the noise is again white and Gaussian and its first and
second moments are given by Eqs.~(\ref{cnoise1})
and~(\ref{cnoise2}) respectively. We note that the sixth order
differential operator has already been deduced in the context of
nonequilibrium growth from atomistic models~\cite{vvedensky2}. The
renormalization group flow equations become
\begin{eqnarray}\label{expnab61}
\frac{d \nu}{d \ell} &=& \nu \left[ z-6-\frac{45}{128 \pi}
\frac{\lambda^2 D}{\nu^3} \right], \\
\frac{dD}{d\ell} &=& D \left[ z-4-2\alpha \right], \label{expnab62} \\
\frac{d \lambda}{d\ell} &=& \lambda \left[ z+\alpha -4 \right].
\label{expnab63}
\end{eqnarray}
We find again the same exponents $\alpha=0$ and $z=4$ due to noise
and vertex non-renormalization (both results are exact again).

The next equation that deserves our attention is a different
variation of Eq.~(\ref{ssgvldsma})
\begin{equation}
\partial_t h = \lambda (h_{xx}h_{yy}-h_{xy}^2) -\nu \nabla^4 h +
\xi(\vec{r},t),
\end{equation}
where the noise is as always white and Gaussian but this time its
mean and correlation are
\begin{eqnarray}\label{mnoisefour}
\langle \xi(\vec{r},t) \rangle &=& 0, \\
\langle \xi(\vec{r},t) \xi(\vec{r} \, ',t') \rangle &=& D \nabla^4
\delta(\vec{r}-\vec{r} \, ') \delta(t-t').\label{cnoisefour}
\end{eqnarray}
This sort of higher order contribution to conserved noise has been
previously considered in the literature~\cite{tang}. In this case,
a simple dimensional analysis shows that the nonlinearity is not
relevant and leads to the exponents $\alpha=-1$ and $z=4$. In this
context a negative exponent $\alpha$ is identified with a flat
interface.

In the case of Eq.~(\ref{sixor})
\begin{equation}
\partial_t h = \lambda (h_{xx}h_{yy}-h_{xy}^2) +\nu \nabla^6 h +
\xi(\vec{r},t),
\end{equation}
where the noise is white and Gaussian and its mean and correlation
are given by Eqs.~(\ref{mnoisefour}) and~(\ref{cnoisefour})
respectively, the RG analysis yields a negative exponent $\alpha$
as well. This can be read from the following RG flow equations:
\begin{eqnarray}
\frac{d \nu}{d \ell} &=& \nu \left[ z-6+\frac{45}{128 \pi}
\frac{\lambda^2
D}{\nu^3} \right], \\
\frac{dD}{d\ell} &=& D \left[ z-6-2\alpha + \frac{3}{512 \pi}
\frac{\lambda^2
D}{\nu^3} \right], \\
\frac{d \lambda}{d\ell} &=& \lambda \left[ z+\alpha -4 \right].
\end{eqnarray}
Thus the exponents become $\alpha=240/179 -2 \approx -0.66$ and
$z=6-240/179 \approx 4.66$. Note that vertex non-renormalization
is again present and exact, while the noise is renormalized at one
loop order. This result, noise renormalization in a conserved
model, was already found in the literature for this kind of noise
and a different drift~\cite{tang}.

Now we direct our analysis to higher order equations like
\begin{equation}
\partial_t h = \lambda (h_{xx}h_{yy}-h_{xy}^2) -\nu \nabla^8 h +
\xi(\vec{r},t),
\end{equation}
where the noise is white and Gaussian with mean and correlation
given by Eqs.~(\ref{cnoise1}) and~(\ref{cnoise2}). The eighth
order differential operator can be considered as a higher order
diffusion mechanism in the context of nonequilibrium growth, when
we regard continuum equations as hydrodynamic descriptions of
atomistic models~\cite{vvedensky2}. This equation can be analyzed
with the RG technique and we have found that neither noise nor
vertex do renormalize at any order, what yields the exact
exponents $\alpha=0$ and $z=4$.

The previous equation
\begin{equation}
\partial_t h = \lambda (h_{xx}h_{yy}-h_{xy}^2) -\nu \nabla^8 h +
\xi(\vec{r},t),
\end{equation}
can be studied with a Gaussian white noise having as first moments
Eqs.~(\ref{mnoisefour}) and~(\ref{cnoisefour}). Its analysis by
means of the RG method yields a negative value of exponent
$\alpha$. In this case the flow equations read
\begin{eqnarray}
\frac{d \nu}{d \ell} &=& \nu \left[ z-8+\frac{1}{\pi}
\frac{\lambda^2
D}{\nu^3} \right], \\
\frac{dD}{d\ell} &=& D \left[ z-6-2\alpha + \frac{3}{128 \pi}
\frac{\lambda^2
D}{\nu^3} \right], \\
\frac{d \lambda}{d\ell} &=& \lambda \left[ z+\alpha -4 \right],
\end{eqnarray}
yielding the exponents $z= 1768/381 \approx 4.64$ and $\alpha =
4-z \approx -0.64$. Again we find that vertex non-renormalization
is exact and noise renormalizes at one loop order.

Our last example is the following equation
\begin{equation}
\partial_t h = \lambda \nabla^4 (h_{xx}h_{yy}-h_{xy}^2) -\nu \nabla^8 h +
\xi(\vec{r},t),
\end{equation}
where the white and Gaussian noise is characterized by the moments
Eq.~(\ref{mnoisefour}) and~(\ref{cnoisefour}). We found this
stochastic differential equation specially interesting because its
RG analysis is meaningless as neither propagator, nor noise, nor
vertex renormalize at any order in this case. So a different
technique should be employed for this kind of model. One
possibility would be the use of self-consistent
expansions~\cite{sce1,sce2,sce3,katzav1,katzav2,katzav3} as we
mentioned in the Introduction, but we have not explored this
possibility yet. We also note that, due to the high order of
differentiation in every term of this equation, its numerical
analysis should be a hard task too.

\section{Origins of scaling}

We now examine the physical origin of scaling relations. The
exactness of the relation $\alpha + z=4$ in all our equations can
be traced back to expansion~(\ref{expansion}). Higher order terms
are meant to describe smaller scale physics. Accordingly, the
exactness of this scaling relation indicates that the nonlinearity
totally dominates over surface diffusion in the large scale.
Considering the full Eq.~(\ref{monge}) with $\mu
>0$ we find, in agreement with Eq.~(\ref{expansion}), that the
Laplacian dominates. So expansion~(\ref{expansion}) provides a
simple geometric interpretation of the renormalization group
results and the exactness of this scaling relation.

The exactness of the exponents for the SSG and VLDS universality
classes, found in our equations, was due to noise and vertex
non-renormalization. Noise non-renormalization is a common fact,
and its origin is in the conserved character of the drift, which
has been shown in previous works~\cite{sun,lai,janssen}. The case
of the vertex is new. Indeed, the vertex structure
\begin{equation}
\label{measure} \lambda \int \frac{1}{2} \left( \vec{k} \cdot
\vec{q}^\perp \right)^2
\end{equation}
implies that all contributions to the renormalization of the
vertex are $O(k^6)$, while the vertex renormalizes at $O(k^4)$.
The concrete calculations have been carried out in
Sec.~\ref{method}. In other words, the vertex is not renormalized
because all the generated Feynman diagrams correspond to
irrelevant contributions. This is easily seen by noting that in
the renormalization group calculation the vertex contributions
result from integrating against measure~(\ref{measure}). The
procedure is identical to the one carried out for the KPZ
equation~\cite{medina} {\it mutatis mutandis}. Any possible
contributions to the renormalized coefficients come from
integrands that are exactly $O(k^0)$, as higher order
contributions are irrelevant. It is immediate realizing that the
$O(k^0)$ contribution of measure~(\ref{measure}) is identically
zero. In consequence, noise and vertex do not renormalize at any
order in the loop expansion. This is in contrast to what happened
to the SGG and VLDS equations, for which the vertex one loop
diagrams were relevant but they cancelled each other
out~\cite{sun,lai}.

While the renormalization group is a perturbative technique which
may allow access to the critical exponents, it is also possible to
unveil (at least part of) the critical behavior by means of
finding the stationary probability distribution of a given
Langevin equation. This approach has being successfully exploited
in the past~\cite{katzav4,katzav5,kardar}, mainly (but not
only~\cite{kardar}) in those cases in which the stationary state
is Gaussian. In our case it is possible as well, as was shown
in~\cite{escudero}, to formally find the exact stationary
probability density corresponding to Eq.~(\ref{ssgvldsma}) using
its gradient flow structure
\begin{equation}
\partial_t h = -\frac{\delta \mathcal{V}}{\delta h}+ \xi(\vec{r},t),
\qquad \mathcal{V}[h]= -\int \left[ \lambda \, h_{x} h_{y} h_{xy}
- \frac{\nu}{2} (\nabla^2 h)^2 \right] dx \, dy.
\end{equation}
Then the probability density reads
\begin{equation}
\mathcal{P} \sim \exp \left\{ \int \left[ \lambda \, h_{x} h_{y}
h_{xy} - \frac{\nu}{2} (\nabla^2 h)^2 \right] dx \, dy \right\},
\end{equation}
and it is evidently not Gaussian. This distribution is of course
not normalizable due to the presence of the cubic term, and
consequently this expression is purely formal. Its form suggests
that the interface profile could be initially dominated by the
quadratic term, and after some transient the noise could drive the
interface out of this potential well. Then for long times the
cubic term would dominate, as the dynamics would be separated from
the initial Gaussian behavior by the nonlinearity. This simple
picture, derived from the formal stationary probability
distribution, agrees with the results of the renormalization group
analysis.

\section{Conclusions}

In this work we have shown that the scaling relation $\alpha +
z=4$, which originates in the vertex non-renormalization and which
is inexact in the renormalization group sense for both SSG and
VLDS equations, becomes exact in the case of Eq.~(\ref{mampere})
with both conserved and non-conserved noises. The physical origin
of this scaling relation is not the existence of a symmetry (such
as a functional Galilean invariance) but the existence of the
geometric principle described by
Eqs.~(\ref{principle})-(\ref{potential})-(\ref{expansion}).

We have proven the exactness of this scaling relation in $d=2$, in
the renormalization group sense, for a series of stochastic
differential equations presented in Sec.~\ref{differentmodels}.
Therein we have found that vertex non-renormalization is exact for
the considered nonlinearity independently of whether noise is
renormalized or not. The validity of these results extends to all
examined models, except the last one. In that case, the higher
order of differentiation precludes the access to the critical
behavior by means of the employment of the RG technique, as we
have shown.

We have focused on conserved surface growth but the existence of
scaling laws happens as well in non-conserved cases. Notably the
KPZ equation obeys the relation $\alpha +z=2$. Although this
scaling relation has been traditionally attributed to the
existence of a Galilean invariance in the equation, recent
numerical work suggests that it is possible to find discrete
schemes which do not obey this symmetry and are still able to
reproduce the KPZ universality class~\cite{wio,wior,wiop}. Perhaps
the existence of a geometric principle, akin to the one discussed
here, could be a possible alternative explanation. The origin of
such a principle is not clear so far, but the variational
formulation of the KPZ equation~\cite{wio2} might have some
relation with it, in case it exists.

\begin{acknowledgments}
The authors are grateful to Rodolfo Cuerno and Ireneo Peral for
useful comments and discussions. This work has been partially
supported by the MICINN (Spain) through Projects No. MTM2010-18128
and No. FIS2009-9870.
\end{acknowledgments}

\appendix

\section{RG Analysis with an asymmetric vertex form}
\label{appasymm}

As mentioned in section~\ref{method} using asymmetric vertex forms
leads to the explicit appearance of the polar angle in Fourier
space. This fact signals the generation of new operators and
requires the employment of a more sophisticated mathematical
treatment that will be outlined at the end of this appendix. The
aim of this appendix is showing that, nevertheless, it is possible
to formally proceed with the calculations ignoring the presence of
this angle and still find the good result. Sometimes this is
possible, sometimes it is not. Herein we exemplify these different
possibilities.

We consider the equation
\begin{equation}
\label{ampere}
\partial_t h = \lambda (h_{xx}h_{yy}-h_{xy}^2)-\nu \nabla^4 h +
\xi^{(1,2)}.
\end{equation}
First of all we shall concentrate on the non-conserved noise case
$\xi=\xi^{(2)}$. We employ the following asymmetric vertex form
\begin{eqnarray}
\nonumber
h(\vec{k},\omega)=h_0(\vec{k},\omega) + \lambda \, G_0(\vec{k},\omega) \times \\
\nonumber \int \left[ (k_x-q_x)^2 q_y^2 - (k_x - q_x)(k_y - q_y)
\, q_x \, q_y
\right] \\
\times \, h (\vec{k}-\vec{q},\omega-\Omega) \, h(\vec{q},\Omega)
\, \frac{d\Omega \, d\vec{q}}{(2 \pi)^3}, \label{fourierampere}
\end{eqnarray}
where
\begin{eqnarray}
h_0(\vec{k},\omega) = G_0(\vec{k},\omega) \, \tilde{\xi}(\vec{k},\omega), \\
G_0(\vec{k},\omega) = (-i \omega +  \nu k^4)^{-1},
\end{eqnarray}
$\tilde{\xi}$ is the Fourier transformed white noise,
$k=|\vec{k}|$ and $\vec{k}=(k_x,k_y)$ (the same holds for
$\vec{q}$). The renormalized coefficients satisfy the RG flow
equations, which at one loop order are given by
\begin{eqnarray}
\frac{d \nu}{d\ell} &=& \nu \left[ z-4-\frac{\lambda^2 D}{8 \pi \nu^3} \cos^2(\phi) \right], \\
\frac{d D}{d\ell} &=& D \left[ z-2-2\alpha \right], \\
\frac{d \lambda}{d\ell} &=& \lambda \left[ z+\alpha -4 \right].
\end{eqnarray}
The variable $\phi$ is the polar angle in Fourier space. An
alternative way of thinking about this angle is through the
relation
\begin{equation}
\cos^2 (\phi)= \frac{\widehat{h_{xx}}}{\widehat{h_{xx}} +
\widehat{h_{yy}}},
\end{equation}
where $\hat{\cdot}$ denotes the spatial Fourier transform. The
values of the exponents are independent of this angle and take the
correct values $\alpha = 2/3$ and $z= 10/3$.

Now we move to the case of conserved noise, i. e. to the model
\begin{equation}
\label{ampere2}
\partial_t h = \lambda (h_{xx}h_{yy}-h_{xy}^2)-\nu \nabla^4 h +
\xi^{(1)}.
\end{equation}
Repeating the dynamic renormalization group calculations we arrive
at the flow equations for the effective parameters
\begin{eqnarray}
\frac{d \nu}{d\ell} &=& \nu \left[ z-4-\frac{\lambda^2 D}{8 \pi \nu^3} \cos^2(\phi) \right], \\
\frac{d D}{d\ell} &=& D \left[ z-4-2\alpha +\frac{D \lambda^2}{64 \pi \nu^3} \right], \\
\frac{d \lambda}{d\ell} &=& \lambda \left[ z+\alpha-4 \right].
\end{eqnarray}
The resulting exponents are independent of $\phi$ and consequently
they are the correct ones: $\alpha = 0$ and $z=4$. Note however
that the noise renormalization that arises in this case should be
interpreted as a spurious result since it does not appear in the
case of a symmetric vertex form, cf.
Eqs~(\ref{rgf11})-(\ref{rgf22})-(\ref{rgf33}).

Now we move to the last model
\begin{equation}
\label{ampere3}
\partial_t h = \lambda (h_{xx}h_{yy}-h_{xy}^2)+\nu \nabla^6 h +
\xi^{(1)}.
\end{equation}
The renormalization group leads us this time to the flow equations
\begin{eqnarray}
\label{theta1}
\frac{d \nu}{d\ell} &=& \\
&=& \nu \left[ z-6 + \frac{\lambda^2 D}{256 \pi \nu^3}
\cos^2(\phi) \{19 - 10 \cos(2 \phi)\} \right],
\nonumber \\
\label{theta2} \frac{d D}{d\ell} &=& D \left[ z-4-2\alpha +\frac{D
\lambda^2}{64 \pi \nu^3} \right], \\
\frac{d \lambda}{d\ell} &=& \lambda \left[ z+\alpha -4 \right].
\label{theta3}
\end{eqnarray}
So we find the $\phi-$dependent exponents
\begin{eqnarray} \label{alfa}
\alpha &=& \frac{8}{-4 + 3 \cos^2(\phi) \{19 - 10 \cos(2 \phi)\}}, \\
z &=& 4 + \frac{8}{4 -3 \cos^2(\phi) \{19 - 10 \cos(2 \phi)\}}.
\label{zeta}
\end{eqnarray}
In this case these values have to be considered invalid because it
is not possible to recover the correct exponents found in the
analysis performed with the symmetric vertex form from
Eqs.~(\ref{alfa}, \ref{zeta}). Indeed, averaging $\phi$ out from
either Eqs.~(\ref{theta1}, \ref{theta2}, \ref{theta3}) or
(\ref{alfa}, \ref{zeta}) does not reproduce the actual values of
the exponents $\alpha=0$ and $z=4$ known form
Eqs.~(\ref{expnab61}, \ref{expnab62}, \ref{expnab63}). The correct
way to proceed now would be deriving the stochastic growth
equation with renormalized coefficients. This leads to the
appearance of new operators in the equation. An example with a
fourth order equation can be found in~\cite{sarma}. In this
example, contrary to ours, a new term in the equation is
generated; in our present case the differential operators become
integro-differential. In fact, $\cos^2(\phi)$ gives rise to a
linear integral operator which in Fourier space reduces itself to
a multiplication by the factor
\begin{equation}
\frac{k_x^2}{k_x^2 + k_y^2}
\end{equation}
of the field over which it is applied~\cite{stein}. The resulting
stochastic growth equation should be analyzed again by means of
the RG technique. Obviously, this is much harder than using a
symmetric vertex form.

In summary, it is possible to analyze stochastic growth equations
using asymmetric vertex forms. One can formally proceed, ignore
the presence of polar angles in Fourier space, and end up with
$\phi-$dependent flow equations. If the resulting exponents are
independent of these angles one can conclude the calculation. If
there is $\phi$ dependence in the exponents, then one has to
derive the equation with the newly generated operators and restart
the analysis. The overall procedure is more complicated than using
a symmetric vertex form.

\end{document}